\documentstyle[twocolumn,prl,aps,epsf]{revtex}

\catcode`@=11
\def\citer{\@ifnextchar [{\@tempswatrue\@citexr}{\@tempswafalse\@citexr[]}}
 
\def\@citexr[#1]#2{\if@filesw\immediate\write\@auxout{\string\citation{#2}}\fi
  \def\@citea{}\@cite{\@for\@citeb:=#2\do
    {\@citea\def\@citea{--\penalty\@m}\@ifundefined
       {b@\@citeb}{{\bf ?}\@warning
       {Citation `\@citeb' on page \thepage \space undefined}}%
\hbox{\csname b@\@citeb\endcsname}}}{#1}}
\catcode`@=12

\def\beq{\begin{equation}}
\def\eeq{\end{equation}}

\def\beqn{\begin{eqnarray}}
\def\eeqn{\end{eqnarray}}
\relax
\def\ba{\begin{array}}
\def\ea{\end{array}}

\def\be{\begin{equation}}
\def\ee{\end{equation}}
\def\bea{\begin{eqnarray}}
\def\eea{\end{eqnarray}}

\def\to{\rightarrow}

\def\f{\frac}
\def\ra{\rightarrow}
\def\hbb{h b \bar b}
\def\zbb{Z b \bar b}
\def\bbbb{b \bar b b \bar b}
\def\bbjj{b \bar b j j}
\def\Psibar{\bar{\Psi}}
\def\bbar{\bar{b}}
\def\tbar{\bar{t}}
\def\thisday{~Phys.~Rev.~Lett., in press, ~and~ hep-ph/9802294~~}

\newcommand{\gae}{\stackrel{>}{\sim}}

\begin{document}                                                              

\title{Higgs Bosons with Large Bottom Yukawa Coupling at Tevatron and LHC
}  
\author{{\sc J.~Lorenzo~Diaz-Cruz}$^1$,~~ {\sc Hong-Jian He}$^2$,~~
        {\sc Tim~Tait}$^{2,3}$,~~ {\sc C.--P.~Yuan}$^2$}
\address{
$^1$~Instituto de Fisica, BUAP, 72570 Puebla, Pue, Mexico \\
$^2$~Michigan State University, East Lansing, Michigan 48824, USA \\
$^3$~Argonne National Laboratory, Illinois 60439, USA
}
\date{\thisday}
\maketitle
\begin{abstract}
\hspace*{-0.35cm}
We study the discovery reach of the Tevatron and the LHC
for detecting a Higgs boson ($h$), predicted in composite models
of the electroweak symmetry breaking or in supersymmetric theories,
with an enhanced $b$-quark Yukawa coupling via
$p\bar{p}/pp \to b\bar{b} h (\to b\bbar )+X$.  Our analysis shows
that studying this process at the Tevatron Run II or the LHC
can provide strong constraints on these models.
\\[0.1cm]
PACS number(s): 12.60.-i, 14.80.Cp, 14.65.Fy, 12.38.Bx
\end{abstract}


\begin{narrowtext}

\vspace*{-0.8cm}
The yet unverified Higgs sector in the standard model (SM)
generates gauge boson masses via spontaneous
electroweak symmetry breaking (EWSB) and 
fermion masses via Yukawa interactions.
The large top quark mass, at the order of the
EWSB scale, suggests that top may play a special
role in the generation of mass.
This occurs in models with dynamical 
top-condensate/topcolor scenarios \cite{review} as well as 
in supersymmetric (SUSY) theories \cite{susy-review}.
Since the
bottom quark is the isospin partner of 
the top quark, its Yukawa 
coupling with a Higgs boson can be closely related to that of the
top quark. As we will show, because of 
the small mass of bottom ($m_b\sim 4.5$\,GeV) relative to 
top ($m_t\sim 175$\,GeV), studying the $b$ Yukawa couplings can
effectively probe new physics beyond the SM.

In this Letter, we study the detection of a Higgs
boson ($h$) at hadron colliders predicted by 
models where bottom has an enhanced Yukawa coupling ($y_b$).
We begin with a model-independent analysis for
Higgs production associated with $b\bar{b}$ jets,
~$p\bar{p}/pp \to b\bar{b} h \to \bbbb +X$, and determine
the discovery reach of 
Tevatron Run II (a $p \bar p$ collider with $\sqrt{S} = 2$\,TeV) and 
LHC (a $p p$ collider with $\sqrt{S} = 14$\,TeV)
for using this production mode to probe
models of dynamical EWSB and SUSY theories.

\vspace*{0.25cm}
\noindent
{\bf 1. Signal and Background}

To perform a model-independent analysis, 
we analyze the signal rate in terms
of $K$, the factor of enhancement over the SM prediction for the
$\hbb \ra \bbbb$ rate.
By definition, $K = {|y_b/(y_b)_{\rm SM}|}^2 {\rm BR}(h \ra b \bar b)
/ {\rm BR}(h \ra b \bar b)_{\rm SM}$,
where $y_b/(y_b)_{\rm SM}$ is the factor enhancing the
$h$-$b$-$\bar b$ coupling relative to the SM coupling,
and BR($h \ra b \bar b$) is the branching ratio.

The relevant backgrounds to this signal at a hadron collider are
QCD production of $\bbbb$ and $\bbjj$, where $j$ denotes a light
quark or gluon, and production of
$\zbb \ra \bbbb$.  We study  the parton level
$\hbb$ and $\zbb$ rates at 
leading order (LO) with the Monte Carlo 
program PAPAGENO \cite{papageno}.
The $\bbbb$ and $\bbjj$ processes are also studied at the
parton level, with matrix elements
obtained from MADGRAPH~\cite{madgraph}.
In all cases, we include the
full matrix elements for the $g g$ as well as $q \bar q$ 
(and in the case of the $\bbjj$ background,
$q g$ and $\bar q g$) initiated
sub-processes, using the CTEQ4L parton distribution functions.
We estimate next-to-leading order
(NLO) effects by including a $k$-factor of 2 \cite{kfactor}
for the signal and all the background rates.
We simulate the detector acceptance by requiring all four of the
final state $b$ (including $\bar b$) quarks to have $p_T \geq 20$\,GeV,
rapidity $|\eta| \leq 2$, and to be separated by a cone of 
$\Delta R \geq 0.4$.

In order to improve the signal to background ratio, we exploit the
typical topology of the $b$ quarks in the signal events.
The $h$ is radiated from one of the primary $b$ quarks (by primary
$b$ quarks, we refer to those which do not result from the $h$
decay), and thus
there is typically one very energetic $b$ quark with momentum on
the order of the mass of $h$, $m_h$, balanced by the
$h$ and the other primary $b$, which is generally much softer.
The $h$ decays into a $b \bar b$ pair with typical transverse
momentum on the order of $m_h / 2$.
Thus, in order to enhance the signal, we order the transverse momenta
of the $b$ quarks, 
{\small $p^{(1)}_T \geq p^{(2)}_T \geq p^{(3)}_T \geq
p^{(4)}_T$}, and require 
{\small $p^{(1)}_T$} $\geq 75$\,GeV, and 
{\small $p^{(2,3)}_T$} $\geq 40$\,GeV.  
These cuts moderately reduce the signal rate by about $23\%$,
while providing a large $90\%$ reduction of the QCD $\bbbb$ background.
(Unless specified, our numerical results shown in this section
are for detecting a scalar of mass $100$\,GeV at the Tevatron.) 

To further improve the signal to background ratio, we
find the pair of $b$ momenta with invariant mass that best reconstructs
$m_h$, and reject the event if the resulting invariant mass is
more than $\Delta m_h$ from $m_h$, where $\Delta m_h$ is
the maximum of either our estimation of the experimental mass resolution 
(which is assumed to be 
$0.10 \; m_h \; \sqrt{ 80{\rm GeV}/ m_h }$ \cite{tev2000} or $10$\,GeV,
whichever is higher)
or the natural width (which is a calculable model-dependent quantity)
of the scalar under study.
We further require that the $p_T$ of the $h$ (taken to be the $p_T$ of
the sum of the 
four-momenta of the two $b$ quarks which best reconstruct
$m_h$) to be $\geq 50$\,GeV,
and that the $b$ quark with largest $p_T$ is not one of
the two $b$ quarks we have associated with the $h$ decay. 
Because the signal
generally produces an $h$ whose $p_T$ must balance 
{\small $p^{(1)}_T$}, this does
not affect the signal rate, but further reduces the QCD
$\bbbb$ background by about $90\%$.

Finally, because the
signal events typically produce a primary $b$ and $\bar b$ on opposite
sides of the detector, we require that the two
$b$'s which were not associated with the $h$ decay have
invariant mass $M_{bb} \geq 50$\,GeV.  This further reduces the
QCD $\bbbb$ background by about $60\%$.
It is interesting to note that despite the fact
that the $g g \ra \hbb$ subprocess composes more than $95\%$ of the
signal rate at the Tevatron if one applies only the basic acceptance
cuts, the situation is entirely different for our optimized
search strategy, for which the $q \bar q \ra \hbb$ process makes up
$93\%$ of the signal.

In order to reduce the huge $\bbjj$ background,
we require all four $b$ quarks to be tagged, and estimate a $60\%$
probability for tagging a $b$ quark satisfying the cuts
described above.  We allow a $0.5\%$ 
probability for a light
quark or gluon to be mistaken for a $b$ quark.  After applying this
requirement, we find that the $\bbjj$ background is very
small.
One can slightly
improve the bounds obtained by requiring only three of the four
jets to be $b$-tagged, thus saving more of the signal rate.

In Tab.~1 we present the resulting number of events at the Tevatron
(assuming 2 ${\rm fb}^{-1}$ of integrated luminosity) for the $\hbb$
signal (assuming $m_h = 100$\,GeV, a mass resolution of 10\,GeV, and
an enhancement of $K = (m_t / m_b)^2 \simeq (39)^2$)
as well as the background processes.  In
the second column we present the number of events after imposing
the minimal acceptance cuts described above.  In the third column,
we show the number of events remaining after employing our
optimized search strategy.  In columns four and five we present
the results after requiring 3 or 4 $b$-tags, respectively.  As
can be seen from the table, our optimized result combined with
3 (4) $b$-tags shows a significance (calculated as the number
of signal events divided by the square root of the number of
background events) of 43.0 (38.3).  Thus it is possible to use
Tevatron data to exclude this choice of parameters at better
than $99\%$.  We obtain similar results for
$m_h$ = 100 (200, 300, 400)\,GeV, finding a total number of background
events (assuming 4 $b$-tags and a Higgs with width less than the
experimental mass resolution) of 59 (4, 2, 0) for a
$10 \; {\rm fb}^{-1}$ Tevatron and 2.5 $\times 10^5$
(4.7 $\times 10^4$, 1.7 $\times 10^4$, 900) for
a $100 \; {\rm fb}^{-1}$ LHC.

\begin{center}
\begin{tabular}{c||c|c|c|c}
\hline\hline
~Process  & Acceptance & Optimized & 3 $b$-tagged & 4 $b$-tagged~ \\ \hline
$\hbb$   & 1690            & 979             & 465  & 127 \\
$\zbb$   & 300             & 54              & 26   & 7   \\
$\bbbb$  & 5071            & 33              & 16   & 4   \\
$\bbjj$  & $5 \times 10^7$ & $2 \times 10^4$ & 73   & 0   \\
\hline\hline
\end{tabular}
\\
\vspace{0.35cm}
\noindent {\footnotesize
\hspace*{-0.2cm}
{\bf Table~1.} The signal and background events for 2~${\rm fb}^{-1}$ 
of Tevatron data, assuming
$m_h = 100$\,GeV, $\Delta m_h = 10$\,GeV, and $K = 39^2$. }
\end{center} 

{}From the results of our study we can determine the minimal value
of $K$, $K_{\min}$,
required to give a $95\%$ C.L. deviation from the background
for a given $m_h$, using
Gaussian statistics when the number of background events is more than
10, and Poisson statistics if it is less than 10.
From this we
determine the curve of $K_{\min}$ versus $m_h$ for the Tevatron Run II
with 2, 10, and 30 ${\rm fb}^{-1}$, and for the LHC with 100
${\rm fb}^{-1}$. Being conservative, we require 4 $b$-tags 
in Secs.~2 and 3 to interpret these results in the context of the
models with dynamical EWSB and the Minimal Supersymmetric SM (MSSM).
   
\vspace*{0.25cm}
\noindent
{\bf 2. Implications for Models of Dynamical EWSB}

Examples of the strongly interacting EWSB sector  
with composite Higgs boson(s) are 
top-condensate and topcolor models \cite{review},
in which new strong dynamics associated with the top quark
play a crucial role for top and ${W,Z}$ mass generation. 
A generic feature of these models is the naturally large
Yukawa coupling of bottom ($y_b$),  
of the same order as that of top ($y_t\sim 1$), due to the
infrared quasi-fixed-point structure \cite{RGE} and their particular
boundary conditions for $(y_b,y_t)$ at the compositeness scale.

\noindent
\underline{\bf Two-Higgs-Doublet Extension of the BHL Model}\\
The effective theory of the top-condensate model 
is the SM without its elementary Higgs boson,
but instead, with 4-Fermi interaction terms induced from 
(unspecified) strong dynamics at a high scale $\Lambda$. 
The minimal BHL top-condensate model with three families \cite{BHL},
contains only one type of 4-Fermi vertex for 
$<\hspace*{-0.15cm}\tbar t\hspace*{-0.15cm}>$ condensation
which generates
the masses for top, $W$, and $Z$, 
but the predicted $m_t$ is too large
to reconcile with experiment. Thus, we consider the two Higgs doublet
extension (2HDE) \cite{tt-2HDM} as an example
(which, with some improvements \cite{review,top-seesaw}, 
is expected to produce an acceptable $m_t$), 
and examine its prediction for the $\hbb$ rate.
The 4-Fermi interactions of the 2HDE model 
produce condensates in both $t\bar{t}$ and $b\bar{b}$ channels, 
which generate the EWSB and 
induce two composite Higgs doublets $\Phi_t$ and $\Phi_b$ so that the
Yukawa interactions take the form of  
$y_t\left(\Psibar_L{\Phi}_tt_R+{\rm h.c.}\right)$  and
$y_b\left(\Psibar_L{\Phi}_bb_R+{\rm h.c.}\right)$. 
Here, $\Psibar_L$ is the left-handed quark doublet and 
$t_R$ is the right-handed top, {\it etc.}
This model predicts 
$~y_t(\Lambda )=y_b(\Lambda ) \gg 1~$ at the scale
$\mu =\Lambda$ \cite{review,tt-2HDM}.
Therefore, 
$~~y_t(\mu )\approx y_b(\mu)$~ for any $\mu <\Lambda$,
because the renormalization group equations governing the running of
$y_t$ and $y_b$ are similar except for the small difference in the
the $t$ and $b$ hypercharges \cite{review,RGE}. Due to the dynamical
$<\hspace*{-0.15cm}\tbar t\hspace*{-0.15cm}>$ and 
$<\hspace*{-0.15cm}\bbar b\hspace*{-0.15cm}>$ condensation, 
the two composite Higgs doublets develop VEVs
{\small $
<\hspace*{-0.15cm}\Phi_t\hspace*{-0.15cm}> 
= \left(v_t, 0\right)^T/\sqrt{2}$} and
{\small $<\hspace*{-0.15cm}\Phi_b\hspace*{-0.15cm}> 
= \left(0, v_b\right)^T/\sqrt{2}$.}
Since $m_b$($=y_bv_b${\small $/\sqrt{2}$})
is fixed by the experimental value at $\mu =m_b$
and $y_b$ is about equal to $y_t(\sim$1), 
this model predicts a large
{\small 
$~~\tan\beta \equiv v_t/v_b\approx m_t/m_b = O(39) \gg 1 ~.$
}

The 2HDE has three neutral scalars, the lightest (with enhanced
$b$ coupling) being the pseudoscalar  
$P$({\small $=\sqrt{2} [\sin\beta {\rm Im}\Phi_b^0
          +\cos\beta {\rm Im}\Phi_t^0]$}),
whose mass ($M_P$) is less than about $233$\,GeV
for $\Lambda =10^{15}$\,GeV \cite{tt-2HDM}.
Given $y_b$ and $M_P$, one can calculate the production rate of
$P b \bar{b}(\rightarrow b {\bar b} b {\bar b})$
at hadron colliders, and thus
for a given $M_P$
one can determine the minimal $y_b$ value needed 
for the Tevatron and LHC to observe the signal.
As shown in Fig.1a,  the Tevatron data  with $2\,{\rm fb}^{-1}$
will exclude such a model at $95\%$C.L.

\noindent
\underline{\bf In Topcolor Assisted Technicolor}\\
The topcolor-assisted 
technicolor models (TCATC) \cite{topcolor} postulate the gauge structure
${\cal G}=
SU(3)_1\otimes SU(3)_2\otimes U(1)_1\otimes U(1)_2\otimes SU(2)_L$ 
at the scale above $\Lambda$ to explain the dynamic origin of the 
4-Fermi coupling(s) in top-condensate models.
At $\Lambda$$\sim$$O$(1)\,TeV, ${\cal G}$ spontaneously breaks down to 
$SU(3)_c\otimes U(1)_Y\otimes SU(2)_L$, and
additional massive gauge bosons are produced in
color octet ($B^a$) and singlet ($Z'$) states. 
Below the scale $\Lambda =\min (M_B,M_{Z'})$, the effective 
4-Fermi interactions are generated in the form of \\[-0.35cm]
{\small 
$$
\begin{array}{l}
{\cal L}_{4F} = \f{4\pi}{\Lambda^2}\left[
\left(\kappa +\f{2\kappa_1}{9N_c}\right)
\Psibar_Lt_R\tbar_R\Psi_L+
\left(\kappa -\f{\kappa_1}{9N_c}\right)
\Psibar_Lb_R\bbar_R\Psi_L \right].  \\[-0.2cm]
\end{array}
$$
} 
\hspace*{-0.35cm}
\noindent\hspace*{-0.25cm}
In the low energy effective theory at the EWSB scale,
two composite Higgs doublets are induced with 
the Yukawa couplings 
$~y_t=\sqrt{4\pi (\kappa +2\kappa_1/9N_c)}~$ 
and
$y_b=\sqrt{4\pi (\kappa -\kappa_1/9N_c)},~$
where $\kappa$ and $\kappa_1$ originate from the strong $SU(3)_1$ 
and $U(1)_1$ dynamics, respectively.  It is clear that, unless $\kappa_1$
is unnaturally larger than $\kappa$,  $y_b$ is expected to be only slightly
below $y_t$.  The $U(1)_1$ force is attractive in the 
$<\hspace*{-0.15cm}\tbar t\hspace*{-0.15cm}>$ 
channel but repulsive in the 
$<\hspace*{-0.15cm}\bbar b\hspace*{-0.15cm}>$
channel, thus $t$, but not $b$, acquires a dynamical mass,
provided
$~~y_b(\Lambda )~<~ y_{\rm crit}=\sqrt{{8\pi^2}/{3}} ~<~ y_t(\Lambda ) ~$.~
(In this model, $b$ acquires a mass 
mainly from the topcolor instanton effect \cite{topcolor}.)
Furthermore, the composite Higgs doublet $\Phi_t$, but not 
$\Phi_b$, develops a VEV, i.e., $v_t\neq 0$ and $~v_b=0$. 
\begin{figure}[htb]
\vspace*{-0.8cm}
\hspace*{-1.5cm}
\epsfxsize=10.3cm\epsfysize=8cm
\epsfbox{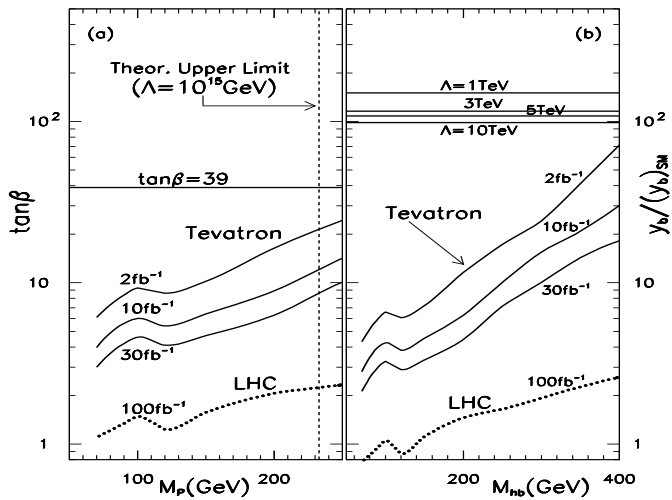}\\[-0.7cm]
\caption[fig:fig2]{Discovery reach of Tevatron and LHC 
for the models of 2HDE (a) and TCATC (b). 
Regions above the curves can be discovered at $95\%$C.L.  
In (b), the straight lines indicate $y_t(\mu =m_t)$ values
for typical values of the topcolor breaking scale $\Lambda$, 
and $y_b$ is predicted to be very close to $y_t$. } 
\end{figure}
In TCATC, the topcolor interaction generates $m_t$, but
is not responsible for the entire EWSB.
Thus, $\Lambda$ can be as low as
$O(1-10)$\,TeV (which avoids the severe fine-tuning needed in the
minimal models \cite{BHL,tt-2HDM}),
and correspondingly, $v_t=64-88$\,GeV for $\Lambda =1-5$\, TeV
by the Pagels-Stokar formula. 
The smaller $v_t$ value (as compared to $v=246$\,GeV)
predicted in the TCATC model 
makes the top coupling to $\Phi_t$ stronger,
i.e., $~y_t=2.8-3.9~$ at $\mu =m_t$,
than in the SM ($y_t \sim 1$).
As explained above, this requires $y_b$ to be
also large.  Thus, the neutral scalars
$h_b$ and $A_b$ in the doublet $\Phi_b$, which are about
degenerate in mass, have an enhanced coupling to the $b$-quark.

In Fig.1b, we show the minimal value of $y_b/(y_b)_{SM}$ needed
to observe the TCATC model signal as a function of $M_{h_b}$. 
As shown, if $M_{h_b}$ is less than about 400\,GeV, the Tevatron
Run II data can effectively probe the scale of the topcolor breaking 
dynamics, assuming the TCATC model signal is observed. 
If the signal is not found, the LHC can further explore this
model up to large $M_{h_b}$.  For example, for
$M_{h_b}=800$\,GeV, the required minimal value of $y_b/(y_b)_{SM}$
is about 4.9 at $95\%$C.L.
Similar conclusions can be drawn for 
a recent left-right symmetric extension \cite{tt-lindner} of the 
top-condensate scenario, which
also predicts a large $b$-quark Yukawa coupling.

\vspace*{0.25cm}
\noindent
{\bf 3. Implications for Supersymmetric Models}

The EWSB sector of the MSSM model includes two Higgs doublets with
a mass
spectrum including two neutral CP-even scalars $h^0$ and $H^0$,
one CP-odd pseudoscalar $A^0$ and a charged pair $H^\pm$.
The Higgs sector is completely determined at tree level by 
fixing two parameters,
conventionally chosen to be $\tan\beta$ 
and the pseudoscalar mass $m_A$ \cite{himssm}.
At loop level, the large radiative corrections to the Higgs boson 
mass spectrum are dominated by the contributions of top and 
stop (top-squark) in loops \cite{himrc}.
In this study we employ the full one loop results \cite{effpot} to
generate the Higgs mass spectrum assuming all sfermion masses, $\mu$,
scalar tri-linear parameters, and SU(2$)_L$ gaugino masses
at the electro-weak scale are equal to 500 GeV.
We find that
our results are fairly insensitive to this choice of parameters.

$\tan\beta$ is a free parameter of the MSSM. The current low energy
bound gives $m_h, m_A >75$\,GeV for $\tan\beta > 1$ \cite{himlim}.
Since the couplings of $h^0$-$b$-$\bar b$, $H^0$-$b$-$\bar b$ and 
$A^0$-$b$-$\bar b$ are proportional to $\sin\alpha/\cos\beta$, 
$\cos\alpha/\cos\beta$ and $\tan\beta$, respectively, they
can receive large enhancing factor when $\tan\beta$ is large. 
This can lead to detectable $\bbbb$ signal events at the Tevatron,
as previously studied in Ref. \cite{daiguve}. Here,
we improve the calculations in the signal and the background rates 
(c.f. Sec.1) and the prediction of the MSSM by including
large loop corrections.
We calculate the enhancement factor $K$
predicted by the MSSM for given values of
$\tan\beta$ and $m_A$. 
In Fig.2 we present the discovery reach of the Tevatron and 
the LHC, assuming a stop mass of 
500\,GeV and that all the superparticles 
are so heavy that Higgs bosons 
will not decay into them at tree level.
The BR for $h\to b\bar b$ is close to one for most of the 
parameter space above the discovery curves. 
Moreover, for $\tan\beta \gg 1$, the $h^0$ is nearly
mass-degenerate with the
$A^0$ (if $m_A$ is less than $\sim$120\,GeV)
and otherwise with $H^0$.
We thus include both scalars in the signal rate provided their
masses differ by less than $\Delta m_h$.
The MSSM can also produce additional $\bbbb$
events through production of $h^0 Z \to \bbbb$ and 
$h^0 A^0 \to \bbbb$, however these rates are expected to be 
relatively small
when the Higgs-bottom coupling is enhanced, and the resulting
kinematics are different from the $\hbb$ signal.  Thus we
conservatively do not include these processes in our signal rate.

From Fig.2 we deduce that if a signal is not found,
the MSSM with $\tan\beta > 23~(16,~12)$
can be excluded for $m_A$ up to 200GeV
at the $95\%$ C.L. by Tevatron data with a luminosity of 
2~(10,~30)~fb$^{-1}$; while the LHC can exclude a much larger
$m_A$ (for $m_A=800$\,GeV, 
the minimal value of $\tan \beta$ is about 7.4). 
These Tevatron bounds thus improve a recent result obtained
by studying the $b\bar{b}\tau\tau$ channel \cite{dressetal}.
We note that studying the $\hbb$ mode can probe an important
region of the $\tan \beta$-$m_A$ plane which is not easily 
covered by other production modes at hadron colliders, such as 
$pp\to t\bar t +h(\to \gamma \gamma)+X$ and
$pp\to h(\to ZZ^*)+X$ \cite{gunorr}.
Also, in this region of parameter space the
SUSY Higgs boson $h^0$ is clearly distinguishable from a SM one.
\begin{figure}[htb]
\vspace*{-0.8cm}
\hspace*{-0.7cm}
\epsfxsize=8.4cm\epsfysize=6cm
\epsfbox{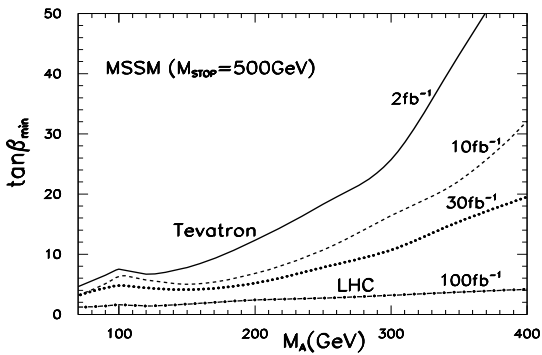}\\[-0.9cm]
\caption{
 The regions above the curves in the $\tan \beta$-$m_A$ plane can
be probed at Tevatron and LHC with a $95\%$ C.L..} 
\label{mssmfig}
\end{figure}

The above results provide a general test for many SUSY models, 
for which the MSSM is the low energy effective theory. 
In the MSSM, the effect of SUSY breaking is parametrized by 
a large set of soft-breaking (SB) terms ($\sim$$O(100)$), 
which in principle should be derived from an underlying model.
We discuss, as examples, the 
Supergravity and Gauge-mediated (GM) models
with large $\tan \beta$.
In the supergravity-inspired model 
\cite{sugrarev} the SUSY breaking occurs in a hidden sector 
at a very large scale, of $O(10^{10-11})$\,GeV, 
and is communicated to the MSSM through gravitational 
interactions.  In the simplest model of this kind, 
all the SB parameters are expressed in terms of 5 universal inputs.
The case of large $\tan\beta$, of $O(10)$, has been examined
within this context \cite{sugratba}, and it was found that
in such a case $m_A$$\sim$$100$~GeV.
Hence, these models can be cleanly confirmed or excluded
by measuring the $\bbbb$ mode at the Tevatron and LHC.

The GM models assume that
the SUSY-breaking scale is much lower, of $O(10^{4-5})$\,GeV, 
and the SUSY breaking is communicated to the MSSM superpartners 
by ordinary gauge interaction \cite{dinetal}.
This scenario can predict large $\tan\beta$ 
($\sim\hspace*{-0.1cm}30$-$40$).  However, in some
models, it favors $m_A \gae 400$\,GeV \cite{bcerom},
which would be difficult to test at the Tevatron, though
quite easy at the LHC.
Nevertheless, in some other models,
a lighter pseudoscalar is possible
(for instance, $\tan\beta=45$ and $m_A=100$) \cite{gmpred},
and the $\bbbb$ mode at hadron colliders can easily explore
such a SUSY model.

In conclusion, the large QCD production rate at a hadron collider 
warrants the detection of a light scalar with large
$h$-$b$-$\bar b$ coupling.
At LEP-II and future $e^+ e^-$ linear colliders, because of the large
phase space suppression factor for producing a direct 3-body final state
as compared to first producing a 2-body resonant state, 
the $b {\bar b} A$ and $b {\bar b} h$ rates predicted by the MSSM are 
dominated by the production of $Ah$ and $hZ$ pairs via 
electroweak interaction. Hence, the $e^+ e^-$ collider is less 
able to directly probe the $h$-$b$-$\bar b$ coupling.

\smallskip
{\small 
We thank E.L. Berger, B.A. Dobrescu, S.~Mrenna, 
C.~Schmidt, R.~Vilar and H. Weerts for useful discussions.
JLDC is supported by CONACYT-NSF international agreement, and 
HJH and CPY by the U.S.~NSF.  TT's work was done at
ANL HEP division and was supported in
part by the U.S. DOE Contract W-31-109-Eng-38.\\[-0.9cm]
}

\end{narrowtext}
\end{document}